\documentclass[pra,aps,twocolumn,nopacs,superscriptaddress,nofootinbib]{revtex4}
\usepackage{graphicx}
\usepackage{dcolumn}
\usepackage{bm}
\usepackage{amsmath}
\usepackage{epsfig}
\usepackage{color}
\usepackage{braket}

\def\jb{{\rm \bf j}} \def\kb{{\rm \bf k}} \def\pb{{\rm \bf p}} \def\rb{{\rm \bf r}} \def\vb{{\bf v}} \def\Fb{{\bf F}}
  \def\qb{{\bf q}}  \def\ub{{\rm \bf u}}  \def\Ub{{\rm \bf U}}
 \def\Eb{\vec{\mathcal{E}}}
\def\vF{{v_{\rm F}}} \def\EF{{E_{\rm  F}}} \def\kF{{k_{\rm F}}}
\def\kB{{k_{\rm B}}} \def\me{{m_{\rm  e}}}
\def\teps{\tilde{E}}  \def\tmu{\tilde{\mu}}
\def\ii{{\rm  i}}   \def\ee{{\rm  e}}
\def\ee{{\rm  e}}   \def\dd{{\rm  d}}
\def\tauim{\tau_{\rm im}} \def\tauac{\tau_{\rm ac}}  \def\tauop{\tau_{\rm op}}

\newcommand\mean[1]{\left<#1\right>}

\begin{document}
\title{Anomalous thermodiffusion of electrons in graphene}
\author{Deng~Pan}
\email[Corresponding author: ]{Deng.Pan@icfo.eu}
\affiliation{ICFO-Institut de Ciencies Fotoniques, The Barcelona Institute of Science and Technology, 08860 Castelldefels (Barcelona), Spain}
\author{Hongxing~Xu}
\affiliation{School of Physics and Technology, Wuhan University, Wuhan 430072, China}
\author{F.~Javier~Garc\'{\i}a~de~Abajo}
\email[Corresponding author: ]{javier.garciadeabajo@nanophotonics.es}
\affiliation{ICFO-Institut de Ciencies Fotoniques, The Barcelona Institute of Science and Technology, 08860 Castelldefels (Barcelona), Spain}
\affiliation{ICREA-Instituci\'o Catalana de Recerca i Estudis Avan\c{c}ats, Passeig Llu\'{\i}s Companys 23, 08010 Barcelona, Spain}


\begin{abstract}
We reveal a dramatic departure of electron thermodiffusion in solids relative to the commonly accepted picture of the ideal free-electron gas model. In particular, we show that the interaction with the lattice and impurities, combined with a strong material dependence of the electron dispersion relation, leads to counterintuitive diffusion behavior, which we identify by comparing a single-layer two-dimensional electron gas (2DEG) and graphene. When subject to a temperature gradient $\nabla T$, thermodiffusion of massless Dirac electrons in graphene exhibits an anomalous behavior with electrons moving along $\nabla T$ and accumulating in hot regions, in contrast to normal electron diffusion in a 2DEG with parabolic dispersion, where net motion against $\nabla T$ is observed, accompanied by electron depletion in hot regions. These findings have fundamentally importance for the understanding of the spatial electron dynamics in emerging material, establishing close relations with other branches of physics dealing with electron systems under nonuniform temperature conditions.
\end{abstract}
\date{\today}

\maketitle

\section{Introduction}

Thermodiffusion, also known as thermophoresis or Soret effect, and widely studied in molecular \cite{DB06,MMA14,WBR10}, and nanoparticle \cite{SWA06,BRH08,JWY09} systems, essentially denotes the fact that a temperature gradient $\nabla T$ in an ensemble of particles can induce a nonuniform density distribution of the constituents. As can be intuitively understood from the equipartition theorem $m\mean{v^2}/2=\xi\kB T$ (in $\xi$ dimensions), particles at higher temperatures move faster and thus should diffuse against the gradient $\nabla T$ towards cold regions, causing depletions of particle density in the hot regions \cite{DB06,JWY09,WBR10}. For conduction electrons in solids, a similar description is also routinely adopted to explain the Seebeck effect, in which a nonzero $\nabla T$ can give rise to a measurable voltage across a material. Although this intuitive prescription correctly describes the signs of the thermoelectric voltage, it does not correctly describe the relation between the Seebeck effect and electron thermodiffusion. As we clarify in this work, electron thermodiffusion is not only driven by the electromotive field in the Seebeck effect, but also by an effective electric field associated with the temperature dependence of the chemical potential. More interestingly, actuated by this total thermoelectric field, electron diffusion could be anomalously directed along the temperature gradient $\nabla T$.

The diffusivity can be quantitatively characterized by a diffusion coefficient $\mathcal{D}=\mean{v^2}\mean{\tau}$, as known from the Fick's laws. For free particles, the relaxation time $\tau$ results only from inter-particle collisions. In contrast, electrons in solids are subject to various relaxation mechanisms, such as scattering by impurities or phonons \cite{HK09}. More importantly, the group velocity of Bloch electrons can have significant departures from free electrons when the dispersion is far from parabolic. A notable example is provided by massless Dirac electrons (MDEs) in graphene \cite{NGM04,CGP09}, which feature a constant group velocity independent of the electron energy. Such a unique conical dispersion could endow graphene with unconventional thermoelectric properties, and indeed anomalous Seebeck properties in graphene materials have already been observed \cite{ZCK09,WBP09,SSX18}. Anticipating a dependence of diffusivity on electron velocity, we can also expect unconventional diffusion behavior of MDEs to emerge in graphene.

In this Letter, we focus on single-layer two-dimensional electron gas (2DEG) and graphene systems for a comparative study of the thermodiffusion behavior associated with either free electrons or MDEs. We find that the free-electron thermodiffusion in 2DEGs behaves indeed in accordance with conventional intuition, whereby electrons at higher temperatures tend to spread to low temperature regions, regardless of the scattering mechanisms. In contrast, MDEs in graphene exhibit an unexpected anomalous thermodiffusion effect. Although a higher temperature in graphene also signifies more energetic electrons, the diffusivity of these electrons is weaker, so that MDEs in graphene diffuse to and accumulate in the hot regions. This result is not only important for thermopower applications, but also it is relevant to ultrafast hot-electron dynamics and its effect on the material response \cite{BTM13,GPM13,WFM14,NWG16,THW18}.

\begin{figure}[t]
\begin{centering}
\includegraphics[width=0.5\textwidth]{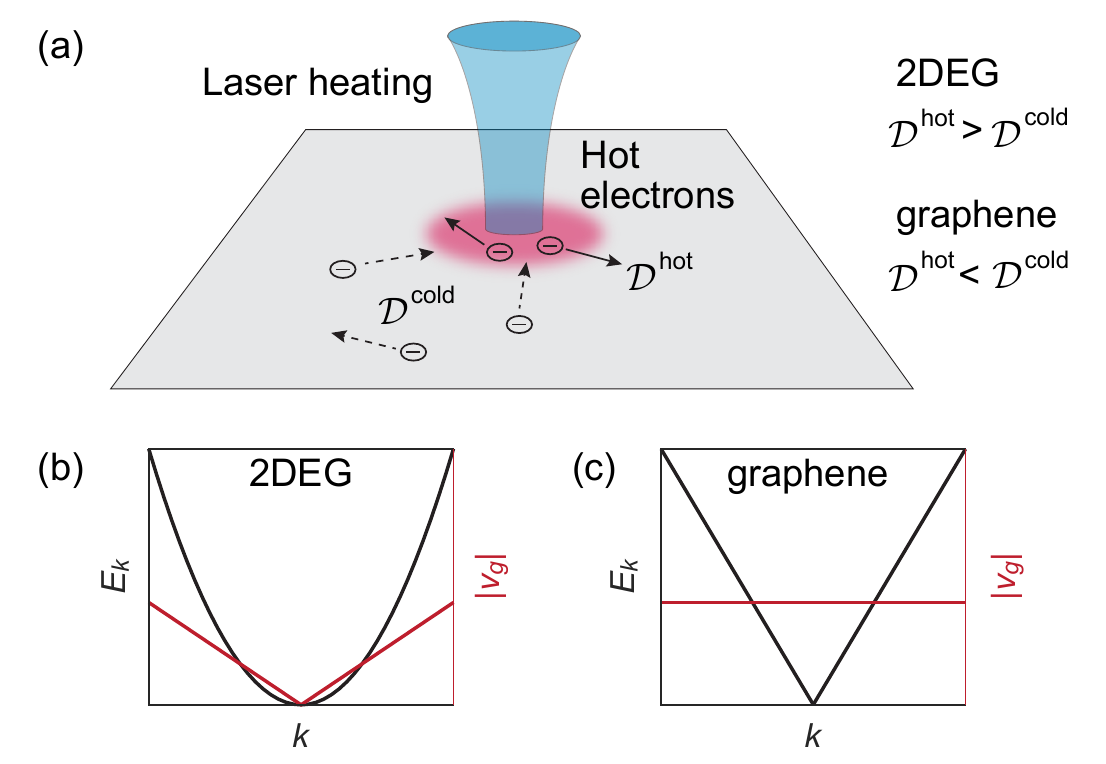}
\par\end{centering}
\caption{ (a) Illustration of a two dimensional system illuminated by a focused CW laser beam to locally heat electrons (red region). (b),(c) Energy dispersion and group velocity of (b) free electrons in a 2DEG and (c) MDEs in graphene. The electron diffusion direction is determined by the temperature dependence of the diffusivity coefficient $\mathcal{D}=\mean{v^2}\mean{\tau}$, which displays the qualitative behavior summarized in (a) (right text), as derived from the electron group velocity in (b) and (c). Normal and anomalous electron diffusion are predicted in a 2DEG and graphene, respectively.
}
\label{Fig1}
\end{figure}

\emph{Qualitative explanation of anomalous thermodiffusion}.--- We consider the thermodiffusion of electrons in a single-layer 2DEG or graphene subject to a temperature gradient. As illustrated in Fig.\ \ref{Fig1}, a strong temperature gradient can be easily achieved by using a tightly focused visible CW laser. The 2DEG can be experimentally realized by a thin doped semiconductor, so in both 2DEG and graphene materials the incident laser can cause strong interband transitions and heat the electrons up to $\sim 10^3\,$K \cite{GPM13,WFM14,NWG16,THW18}. With a temperature gradient established, electron thermodiffusion occurs because the diffusivity characterized by the coefficient $\mathcal{D}=\mean{v^2}\mean{\tau}$ varies with spatial location. For a 2DEG with parabolic dispersion $E_k=\hbar^2k^2/2\me$ [black curve in Fig.\ \ref{Fig1}(b), assuming a bare electron mass $\me$ throughout the paper], the group velocity $\sqrt{2 E_k/\me}$ is proportion to the wave vector $k$, so hot regions containing more energetic electrons should show stronger diffusivity (i.e., $\mathcal{D}^{\rm hot} > \mathcal{D}^{\rm cold}$), if we assume a constant relaxation time $\tau$. In fact, even when taking into account the energy dependence of the relaxation time $\tau(E)$, higher temperature still yields larger $\mathcal{D}$ in the 2DEG (see below). In consequence, the thermodiffusion direction of free electrons is consistent with particles in the ideal gas model.

In stark contrast to the 2DEG, MDEs in graphene exhibit anomalous thermodiffusion with net motions along the temperature gradient as a result of the constancy of the group velocity $\vF\approx 10^8\,$cm/s [Fig.\ \ref{Fig1}(c)]. Assuming a constant $\tau$, the diffusivity coefficient $\mathcal{D}$ is independent of temperature, therefore producing no thermodiffusion of MDEs. In practice, we need to consider an energy dependence of $\tau(E)$ resulting from various scattering mechanisms, so in general, electrons with higher energy possess shorter relaxation time, as shown below based on impurity and phonon scattering calculations. MDEs therefore show stronger diffusivity in hot regions (i.e., $\mathcal{D}^{\rm hot}<\mathcal{D}^{\rm cold}$), leading to anomalous diffusions, with electrons accumulated in the hot areas.

\begin{figure}[t]
\begin{centering}
\includegraphics[width=0.5\textwidth]{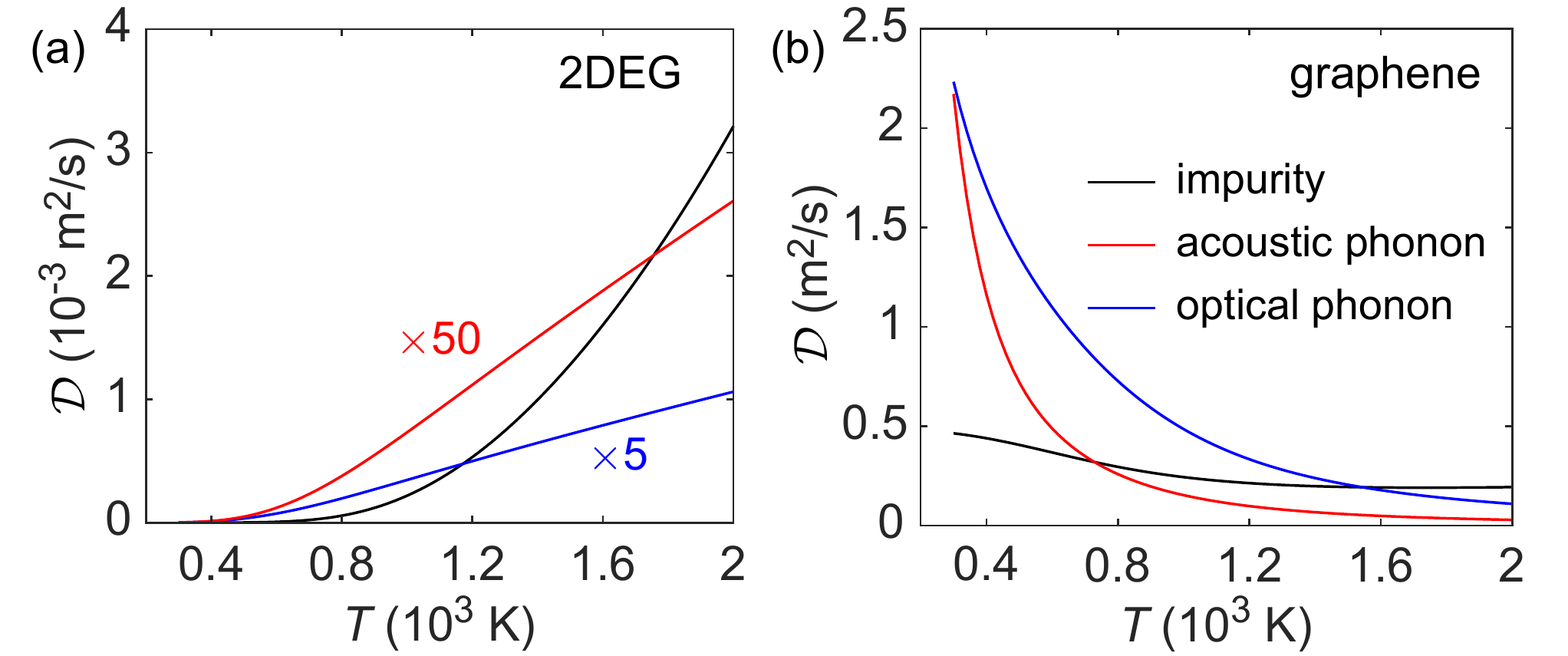}
\par\end{centering}
\caption{ Electron diffusion coefficient $\mathcal{D}$ in a (a) 2DEG and (b) graphene as a function of temperature. Each curve only considers one scattering mechanism caused by impurities (black), acoustic phonons (blue), or optical phonons (red). The temperature dependence of $\mathcal{D}$ supports the conclusion anticipated in Fig.\ \ref{Fig1}(a). Fermi level and impurity density are assumed to be $E_{\rm F}=0.15\,$eV and $n_{\rm i}=2\times 10^{11}\,\rm cm^{-2}$ in both materials (see main text for other parameters).
}
\label{Fig2}
\end{figure}

\begin{figure*}[t]
\begin{centering}
\includegraphics[width=1\textwidth]{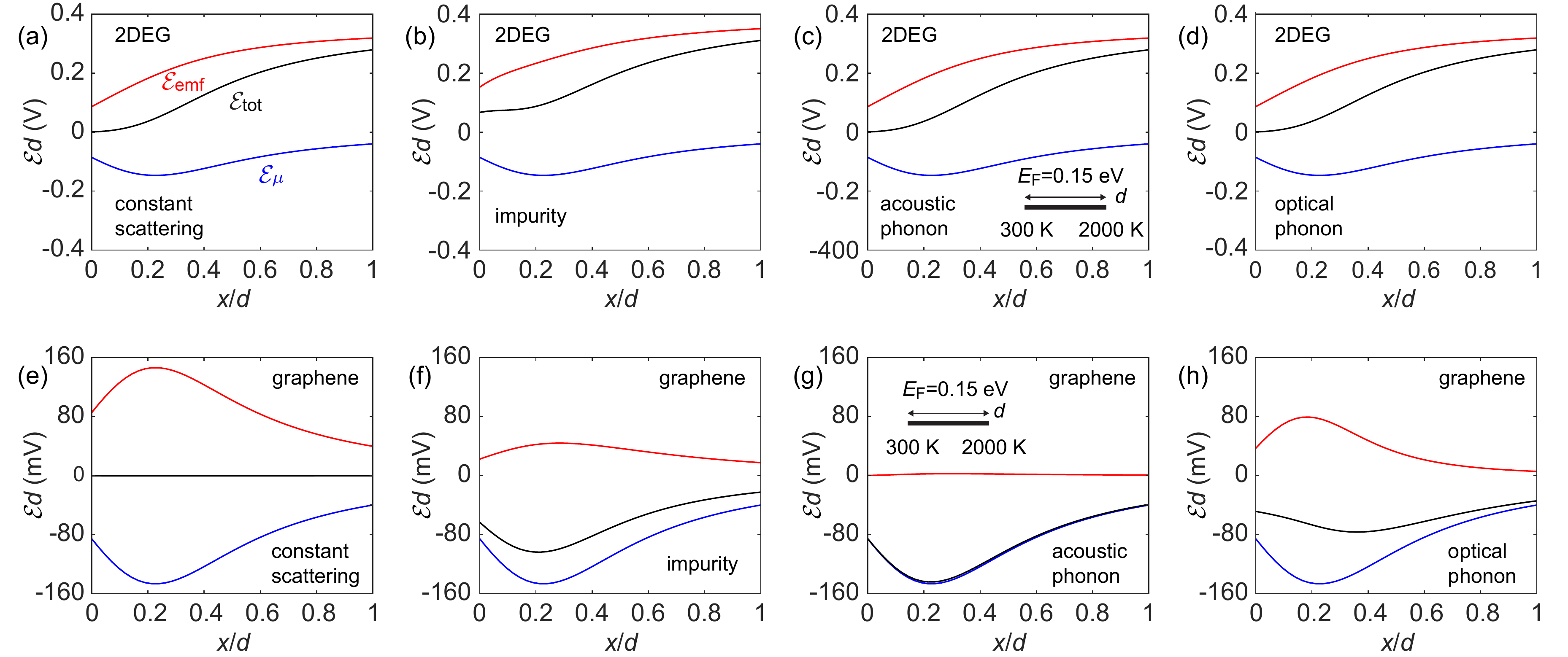}
\par\end{centering}
\caption{ Thermoelectric field acting on a uniform (a)-(d) 2DEG and (e)-(f) graphene subject to a temperature gradient. We consider different models for scattering: (a),(e) a constant scattering rate, (b),(f) impurity scattering, (c),(g) acoustic-phonon scattering, and (d),(h) optical-phonon scattering. A uniform temperature gradient is assumed over a length of $d$: $T(x)=1700[{\rm K}]x/d+300[{\rm K}]$ [see insets in (c),(g)]. Electron diffusion is determined by the total thermoelectric field $\Eb_{\rm tot}$ (black curves), which is the sum of the electromotive field associated with the Seebeck effect $\Eb_{\rm emf}$ (red curves) and an effective field $\Eb_\mu$ (blue curves), arising from the temperature dependence of the chemical potential. The force $-e\Eb_{\rm tot}$ drives electron diffusion opposite and along the temperature gradient in a 2DEG and graphene, respectively, for all scattering mechanisms considered (see signs of $\Eb_{\rm tot}$), except when assuming an unrealistic constant relaxation time in graphene [see (e)], which leads to $\Eb_{\rm tot}=0$. All parameters are the same as in Fig.\ \ref{Fig2}.
}
\label{Fig3}
\end{figure*}

\section{Diffusion coefficients and energy-dependent relaxation}

Although it is not a rigourous procedure, we first estimate the temperature dependence of the electrons using the conventional definition of the diffusion coefficient $\mathcal{D}=\mean{v^2}\mean{\tau}$. A more rigorous model is presented in next section. The mean square velocity of 2DEG electrons is determined by the their thermal energy $\me \mean{v^2}/2=\mean{E_T}-\mean{E_{T=0}}$ at temperature $T$, and for MDEs in graphene one trivially gets $\mean{v^2}=v_{\rm F}^2$.

Both the definition of $\mathcal{D}$ and the more rigourous model used below requires information on the energy dependence of the relaxation time $\tau(E)$. In $\mathcal{D}$, the relaxation time enters through the average
$\mean{\tau}= \int N(E)\tau(E) [-\partial_E f(E)]    /\int N(E) [-\partial_E f(E)]     ,
$
where the density of states is $N(E)=m^2/\pi\hbar^2$ for 2DEG and $N(E)=2|E|/\pi\vF\hbar^2$ for graphene, and $f(E)$ is the Fermi-Dirac distribution, involving a temperature-dependent chemical potential $\mu(T)$. Throughout this paper, the electron energy integral runs over $\{0,+\infty\}$ for 2DEG and $\{-\infty,+\infty\}$ for graphene. The relaxation time is taken to satisfy $\tau(-E)=\tau(E)$.

In this study, we consider energy-dependent relaxations associated with three major processes, corresponding to scattering by impurities, acoustic phonons and optical phonons. Assuming ionic impurities located on the plane of the 2D material and each of them has a charge $e$, the resulting scattering rate is determined by \cite{A06_2,HAD07}
\begin{align}
\frac{1}{\tauim(E_k,T)}=\frac{2\pi n_{\rm i}}{\hbar}\sum_{\ell \kb'} \left|\frac{v}{\epsilon}\right|^2 F[1-\cos(\theta_{\kb'}-\theta_{\kb})] \delta_{k,k'},  \nonumber
\end{align}
where $n_{\rm i}$ is the density of impurities, $\ell$ runs over the two $\pi$ bands in graphene with energy $E_k=\pm\vF\hbar |k|$, $\ell=1$ for 2DEG, $v=2\pi e^2/q$ is the 2D Fourier component of Coulomb potential with wave vector $\bf{q}=\kb'-\kb$, $F=1$ for 2DEG, and $F=[1+\cos(\theta_{\kb'}-\theta_{\kb})]/2$ for graphene. The temperature dependence is incorporated in the screening function $\epsilon(q,T)=1+v\chi(q,T)$, where $\chi(q,T)$ is the susceptibility \cite{AFS1982,WSS06,HD07}. Scattering by this type of impurity is dominant in graphene at low and room temperatures \cite{HAD07}.

The scattering rate of electrons by acoustic phonons can be written as \cite{HD08,SCP14}
\begin{align}
\frac{1}{\tauac(E,T)}= \alpha \frac{\pi D_{\rm ac}^2}{4\hbar\rho v_{\rm ac}^2} \kB T N(E),    \nonumber
\end{align}
and for optical phonons the rate is given by \cite{SK12,SCP14,HAD15}
\begin{align}
\frac{1}{\tauop(E,T)}= \alpha \frac{\pi D_{\rm op}^2}{8\rho \omega_0} \sum_\pm \left[n_T(\omega_0) + \frac{1}{2}\pm\frac{1}{2}\right] N(E\mp\hbar\omega_0),  \nonumber
\end{align}
where $\alpha=4$ for 2DEG and $\alpha=1$ for graphene, $D_{\rm ac}$ and $D_{\rm op}$ are the deformation potentials of acoustic and optical phonons, $\rho$ is the area mass density, $\hbar\omega_0$ is the optical phonon energy, and $n_T(\omega_0)$ is the Bose-Einstein distribution. We note that $n_T$ should be evaluated at the lattice temperature, which can be much lower than the electron temperature \cite{GPM13}. For convenience, we assume equal temperatures of electrons and lattice at each spatial location, a condition that does not affect the main results of this work. We also use the effective acoustic velocity $v_{\rm ac}$, defined by $2/v_{\rm ac}^2=1/v_{\rm ac,L}^2+1/v_{\rm ac, T}^2$ and accounting for both longitudinal and transverse phonons \cite{SCP14}.

With the energy and temperature dependent $\tau(E,T)$ obtained from the above equations, we can readily find the diffusion coefficient $\mathcal{D}$ for electrons at different temperatures, as shown in Fig.\ \ref{Fig2}. Here, for both we use the same parameters $D_{\rm ac}=9.94\,$eV, $D_{\rm op}=5\times 10^9\,$eV/cm, $\hbar\omega_0=147\,$meV, $\rho=7.6\times 10^{-8}\,\rm g/cm^2$, and $v_{\rm ac}=1.62\times 10^6$cm/s, which are in fact chosen to be consistent with the properties of graphene \cite{HS08,SCP14,HAD15}. Actually, results presented below on the thermoelectric field do not strongly depend on these parameters. In Fig.\ \ref{Fig2}, for all the three scattering mechanisms considered, $\mathcal{D}$ increases monotonically with temperature in the 2DEG, while it decreases for MDEs in graphene at high temperatures, thus confirming normal and anomalous diffusion regime in the 2DEG and graphene, respectively.

\section{Model based on the Boltzmann equation}

A standard and more rigorous description of electron diffusion is based on Boltzmann's transport equation, which is valid for small temperature gradients as those here considered. According to this model, the total electric field acting on the electrons can be written
\begin{align}
\Eb_{\rm tot}=  \Eb_{\rm Coul}+ \frac{1}{e} \nabla \mu -S \nabla T,   \label{Etot}
\end{align}
where $\Eb_{\rm Coul}$ is the Coulomb electric field produced by the nonuniform electron charges, the second term $\Eb_\mu=\nabla \mu/e=\partial_T\mu\nabla T/e$ is due to the temperature dependence of the chemical potential, and $\Eb_{\rm emf}=-S \nabla T$ is the electromotive field in the Seebeck effect. It should be noted that for most studies on thermoelectricity, the field component $\Eb_\mu$ is combined with $\Eb_{\rm Coul}$ to define the electrochemical potential that corresponds to the voltage measurable through an external electric circuit, as observed in thermocouples. However, for closed systems such as in Fig.\ \ref{Fig1}(a), the direction of electron diffusion depends on the details of both $\Eb_\mu$ and $\Eb_{\rm emf}$. In particular, the Seebeck coefficient is given by $S= - \mathcal{J}_1/\mathcal{J}_0 eT $ in terms of
$
\mathcal{J}_m= (\pi\hbar^2)^{-1}\int |E|\dd E \tau(E) (E-\mu)^m [-\partial_E f(E)]
$.

\begin{figure}[t]
\begin{centering}
\includegraphics[width=0.5\textwidth]{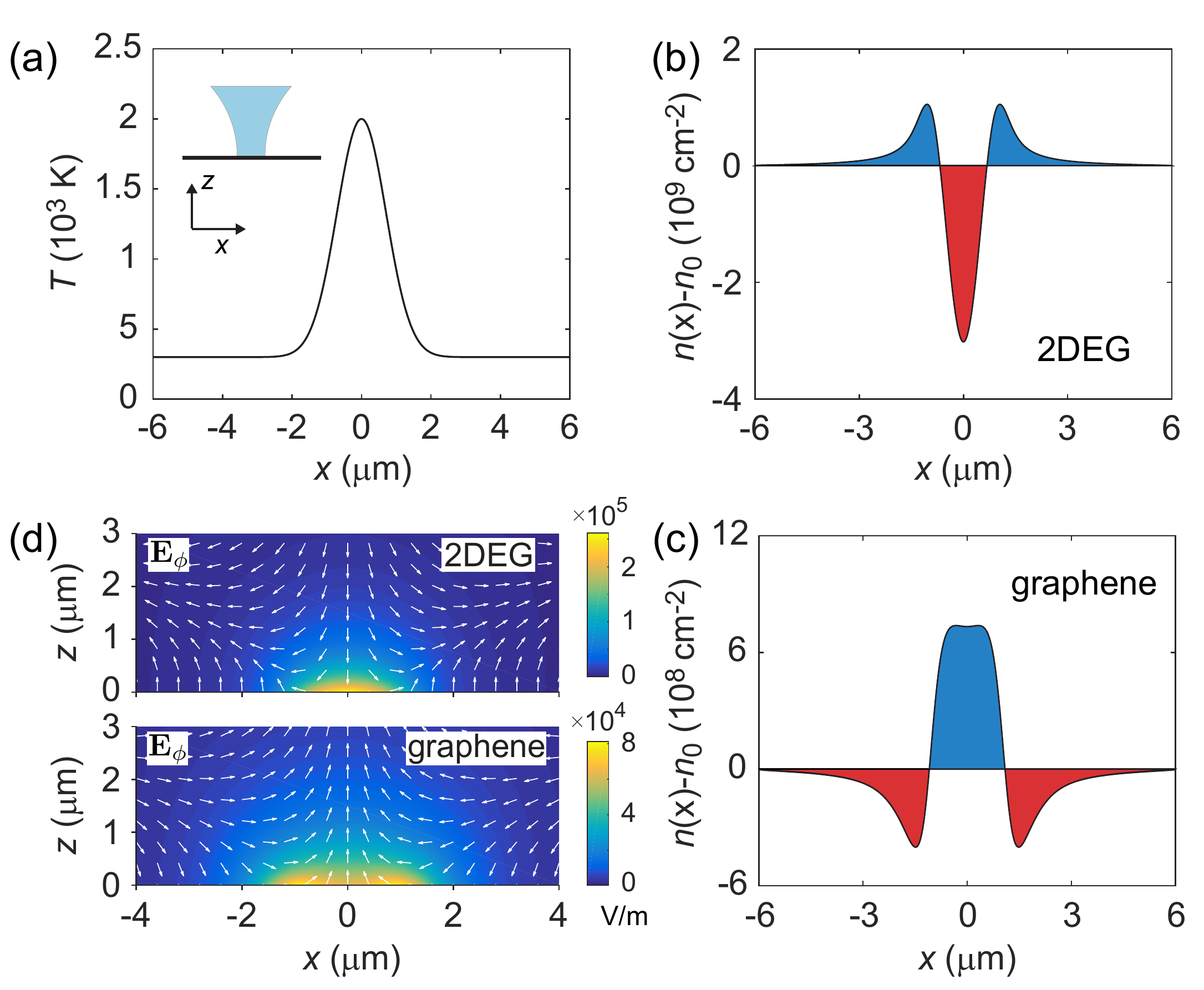}
\par\end{centering}
\caption{ (a) Illustration of a Gaussian temperature distribution in a 2D system: $T(x)=1700[{\rm K}] \ee^{-x^2/w^2} +300[{\rm K}]$ with $w=2\,\mu$m, produced by focused laser heating (inset). (b)-(c) Steady-state electron density in a (b) 2DEG and (c) graphene under the temperature distribution in (a). Here, $n_0$ is the unperturbed electron density, determined by the Fermi level $E_{\rm F}=0.15\,$eV. (d) Amplitude (color scale) and orientation (arrows) distributions of the static electric fields generated by the thermally excited nonuniform electron charges [color in (b),(c)]. The three scattering mechanisms by impurities and phonons shown in Fig.\ \ref{Fig3} are all included using the same calculation parameters.
}
\label{Fig4}
\end{figure}

\section{Thermoelectric field acting on the material electrons}

In Fig.\ \ref{Fig3}, we show calculations of the thermoelectric field acting on electrons in 2DEG [Fig.\ \ref{Fig3}(a)-(d)] and graphene [Fig.\ \ref{Fig3}(e)-(h)] materials, assuming a constant temperature gradient [see insets in Fig.\ \ref{Fig3}(c),(g)] and uniform density (i.e., $\Eb_{\rm Coul}=0$). We study a region of length $d$ with a uniform temperature gradient between $300\,$K and $2000\,$K. To disentangle the contributions of different relaxation mechanisms, we only consider the energy-dependent relaxation time $\tau(\epsilon)$ through a single scattering channel in each plot of Fig.\ \ref{Fig3}. In addition to scattering associated with impurities [Fig.\ \ref{Fig3}(b),(f)] and phonons [Fig.\ \ref{Fig3}(c),(d) and (g),(h)], as investigated in Fig.\ \ref{Fig2} , we also include here results obtained for a constant relaxation time $\tau_{\rm const}$ [Fig.\ \ref{Fig3}(a),(e)]. Noticing that $\Eb_\mu$ is a material property, and also that both a constant prefactor in $\tau(E)$ or the chosen $\tau_{\rm const}$ cancel out in the above definition of the Seebeck coefficient $S$, we conclude that the thermoelectric field is independent of the actual magnitude of $\tau$.

Due to the vanishing of $\Eb_{\rm Coul}$ for homogeneous electron distributions, the total field reduces to $\Eb_{\rm tot}=\Eb_{\rm emf}+\Eb_{\mu}$, so the electron diffusion direction is determined by the relative magnitude of $\Eb_{\rm emf}$ and $\Eb_{\mu}$. In all calculations, the electromotive force $-e\Eb_{\rm emf}$ [red curves, Fig.\ \ref{Fig3}] is always directed opposite the temperature gradient $\nabla T$ due to the negative Seebeck coefficient in all electron doped systems. In contrast, $-e\Eb_\mu$ is always along $\nabla T$ because an increase in electron temperature lowers the chemical potential $\mu$ (see Eq.\ (\ref{Etot})).

For a 2DEG [Fig.\ \ref{Fig3}(a)-(d)], all scattering mechanisms result in a similarly positive thermoelectric field (black curves), which confirms that free electrons undergo conventional diffusions. In contrast, for a graphene layer with realistic scattering processes due to impurities and phonons [Fig.\ \ref{Fig3}(f)-(h)], $\Eb_\mu$ dominates over $\Eb_{\rm emf}$, which results in a total force $-e\Eb_{\rm tot}$ along $\nabla T$, thus producing anomalous electron diffusions of MDEs. Interestingly, although a constant relaxation time [Fig.\ \ref{Fig3}(b)] can give rise to a nonzero Seebeck effect, $\Eb_{\rm emf}$ is perfectly cancelled by $\Eb_\mu$, which confirms our intuitive explanation in Fig.\ \ref{Fig1}.

\emph{Charge rearrangement in a temperature gradient}.--- The thermoelectric fields $\Eb_{\rm emf}$ and $\Eb_{\rm \mu}$ revealed for uniform electrons in Fig.\ \ref{Fig3} further cause electron diffusions, until a steady electron density distribution is established, when $\Eb_{\rm emf}+\Eb_{\rm \mu}$ is balanced by the additional $\Eb_{\rm Coul}$ produced by the nonuniform charge distribution.

Here, we show the resulting electron rearrangement for a Gaussian distribution of temperature [profile in Fig.\ \ref{Fig4}(a)] in a 2DEG and graphene [Fig.\ \ref{Fig4}(b),(c), respectively]. The thermoelectric field $\Eb_{\rm emf}$ and $\Eb_{\rm \mu}$ induced by the temperature gradient can be directly calculated following a similar approach as used for the calculations of Fig.\ \ref{Fig3}, where we can safely assume a constant $E_{\rm F}$ because the density rearrangement is relatively small. The electric field induced by the nonuniform charge satisfies the Poisson equation $\nabla\cdot\Eb_{\rm Coul}=-4\pi e n(\rb)$, and together with the equilibrium condition $\Eb_{\rm Coul}=-\Eb_{\rm emf}-\Eb_{\rm \mu}$, we can easily find the steady charge distributions in the 2DEG [Fig.\ \ref{Fig4}(b)] and graphene [Fig.\ \ref{Fig4}(c)]. The highest temperature gradient around the waist of the Gaussian distribution results in strong outward and inward electron drifts in the normal and anomalous diffusion regimes found in the 2DEG and graphene, respectively. Such electron drifts further decrease or increase the electron density in the temperature Gaussian peak (around $x=0$), by accumulating or depleting electrons outside the waist. Additionally, nonuniform charge distributions [colors in Fig.\ \ref{Fig4}(b),(c)] associated with the electron density rearrangements can induce static electric fields $\Eb_{\rm Coul}$ to balance the thermoelectric field. The spatial distributions of $\Eb_{\rm Coul}$ for the 2DEG and graphene outside the film are shown in Fig.\ \ref{Fig4}(d). The field $\Eb_{\rm Coul}$ is localized around the span of the temperature Gaussian distribution, with the field directions determined by the signs of charges accumulated by electron thermodiffusions in Fig.\ \ref{Fig4}(b),(c).

\section{Concluding remarks}

In summary, we have shown that electron thermodiffusion in solids portrays richer physics than in free particle systems, due to a wealth of scattering mechanisms and ways in which the electron velocity depends on electron energy, reflecting the wide variety of electron band structures near the Fermi level. We expect anomalous thermodiffusion to be found in MDEs in various 2D or higher-dimensional materials, such as surface states of topological insulators \cite{KM05,KWB07,HK07,HQW08,DOL13} or Dirac semi-metals \cite{YZT12,LZZ14_2}. The concepts of conventional and anomalous thermodiffusion can be straightforwardly generalized to holes in n-doped semiconductors and graphene. In addition, the Seebeck coefficients of electrons in some metals (e.g., nickel and potassium) are intrinsically positive. We expect that anomalous electron thermodiffusion will also be found in such materials.

Our findings are important for a fundamental understandings of the evolution of electrons in a material subject to a temperature gradient, which can be relevant to applications in thermopower generation. The processes of electron thermodiffusion here revealed can be related to branches of physics dealing with electron systems in nonuniform temperature environments. For example, when further considering the dynamical establishment of a nonuniform charge density as shown in Fig.\ \ref{Fig4}(b),(c), a focused laser pulse can directly excite charge oscillations in extended graphene, thus offering a sought-after way to generate graphene plasmons in extended homogeneous layers without resorting to scattering structures \cite{CHC16,SM17,paper313}, such as the tips commonly used in scanning near-field optical microscopy \cite{paper196,FRA12,LGA17}. The interaction between the electric field induced by the thermally excited charges and neighboring molecules could lead to a new way of performing nonlinear optical sensing \cite{paper256,paper279,paper315}. The electric potential built up by a nonuniform electron charge distribution as shown in Fig.\ \ref{Fig4}(d) could interact with an electron beam and thus provide the a potential way to realize a phase plate for ultrafast electron beam shaping \cite{paper312,paper332,KG20}.

\appendix

\section{Thermal properties of two dimensional systems}
\label{secS1}

\subsection{Temperature-dependent chemical potential}

\begin{figure*}
\begin{centering}
\includegraphics[width=1\textwidth]{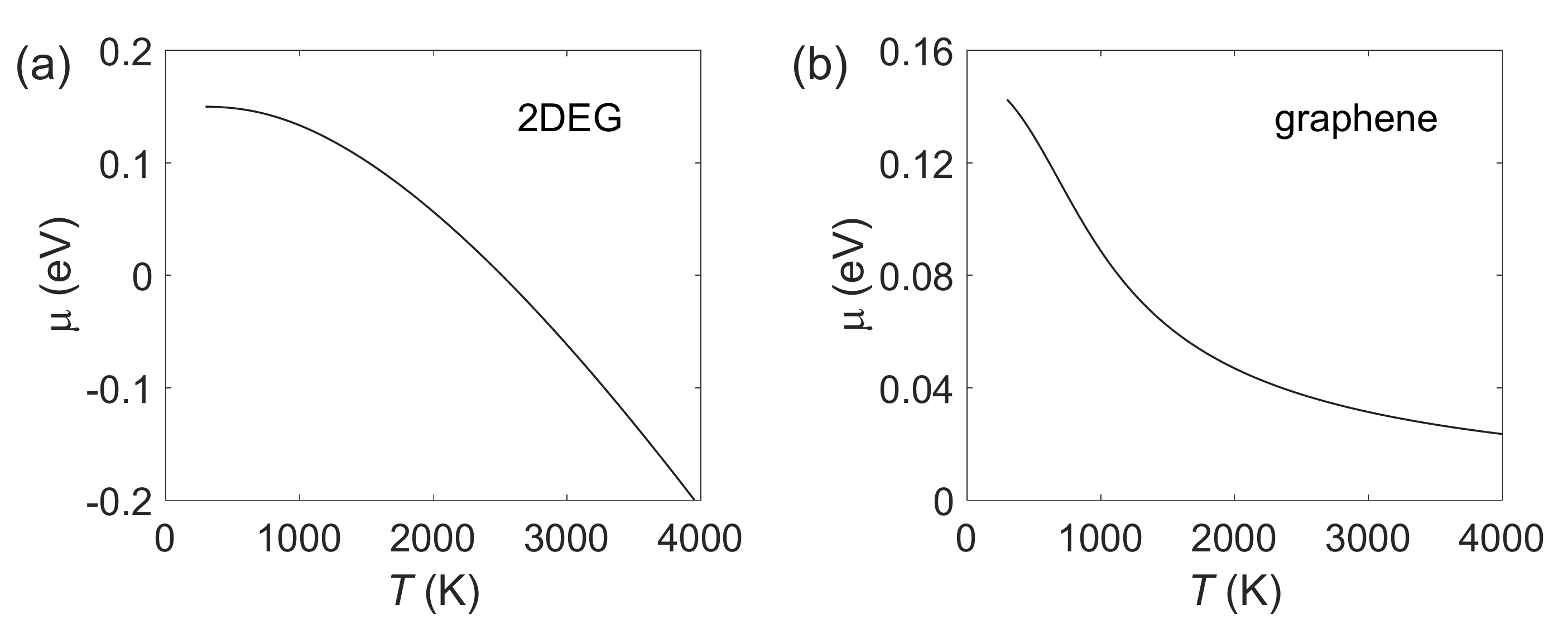}
\par\end{centering}
\caption{Temperature dependence of the chemical potentials in a 2DEG (a) and graphene (b). The Fermi energy is assumed to be $\EF=0.15\,$eV in both cases.}
\label{FigS1}
\end{figure*}

For a homogeneous two-dimensional electron system at finite temperature $T$, the charge density is determined by
\begin{align}
n_{\rm e} &= \int_0^{+\infty}  N(E) \dd E  \frac{1}{\ee^{(E-\mu)/\kB T}+1}        \nonumber\\
          &+ \int_{-\infty}^0  N(E) \dd E  \left(1-\frac{1}{\ee^{(E-\mu)/\kB T}+1}\right)    ,\label{mu}
\end{align}
where $\Theta(x)$ is the step function and we use the density of states $N(E)=\Theta(E) \me/\pi\hbar^2 $ for a 2DEG and $N(E)=2|E|/\pi\vF^2\hbar^2$ for graphene. We consider only one parabolic band for the 2DEG with energy dispersion $E=\hbar^2 k^2/2\me$, and two conical bands in graphene with energies $E=\pm \vF \hbar |k|$. Assuming the material to be doped to a Fermi energy $E_{\rm F}$, the Fermi-Dirac distribution reduces to $\Theta(E_{\rm F}-E)$  at zero temperature $T=0$, so Eq.\ (\ref{mu}) relates the electron density $n_{\rm e}$ to $E_{\rm F}$ through $n_{\rm e}=\me E_{\rm F}/\pi\hbar^2$ for the 2DEG and $n_{\rm e}= E_{\rm F}^2/\pi\hbar^2 v_{\rm F}^2$ for graphene. Inserting these relations back into Eq.\ (\ref{mu}), we find
\begin{align}
\tilde{E}_{\rm F} = \int_0^\infty  \dd \teps \frac{1}{\ee^{\teps-\tmu}+1}            \label{mu2DEG}
\end{align}
for the 2DEG, and
\begin{align}
\tilde{E}_{\rm F}^2=2\int_0^\infty \teps \dd \teps     \left( \frac{1}{\ee^{\teps-\tmu}+1}  -    \frac{1}{\ee^{\teps+\tmu}+1}\right)    \label{mugrap}
\end{align}
for graphene, where a tilde is used for normalized quantities such as $\tilde{E}_{\rm F}=\EF/\kB T$, $\teps=E/\kB T$, and $\tmu=\mu/\kB T$. The solutions of these equations give the temperature dependence of the chemical potential in both types of materials. Numerical solution of Eqs.\ (\ref{mu2DEG}) and (\ref{mugrap}) yields the temperature-dependent chemical potentials shown in Fig.\ \ref{FigS1}. We find that $\mu$ drops below zero with increasing temperature in the 2DEG, while it asymptotically approaches zero in graphene.

\subsection{Electron thermal energy}

In the main text, we estimate the mean square velocity $\mean{v^2}$ from the thermal energy instead of the total electron energy because from the semiclassical perspective the diffusion coefficient $\mathcal{D}$ should be zero at $T=0$ due to freezing of diffusion motion. The average electron energy at finite temperature is given by
\begin{align}
\mean{E_T}= \int  N(E) E \dd E  \frac{1}{\ee^{(E-\mu)/\kB T}+1} .\nonumber
\end{align}
The integration limits are here understood to run over $\{0,+\infty\}$ for the 2DEG and $\{-\infty,+\infty\}$ for graphene. The average electron energy at zero temperature $\mean{E_{T=0}}$ can be analytically found by writing the Fermi-Dirac distribution as $\Theta(E_{\rm F}-E)$. In our study, we calculate the average electron energy at finite temperature numerically, and then estimate the mean square velocity in the 2DEG by using $\me \mean{v^2}/2=\mean{E_T}-\mean{E_{T=0}}$. For MDEs in graphene, one trivially has $\mean{v^2}=v_{\rm F}^2$.

\section{Thermoelectric field described through the Boltzmann equation}

We discuss here a description of electron thermodiffusion based on Boltzmann's transport equation (BTE), where the macroscopic electron properties can be statistically captured by the distribution function $f(\rb,\pb,t)$ as a function of space, electron momentum, and time. Since we are only interested in the quasi-equilibrium limit, the electrons at each spatial location $\rb$ can be characterized by a well-defined temperature $T(\rb)$, following the Fermi-Dirac distribution $f(E)=1/[\ee^{(E-\mu)/\kB T}+1]$, where $\mu(\rb,t,T)$ is the temperature-dependent local chemical potential, and the electron energy $E$ is related to the momentum $\pb$ by the electron dispersion relation, which depends on the choice of material.

According to Eq.\ (\ref{mu}), the chemical potential $\mu(\rb,t,T)$ is determined by the local electron density $n_{\rm e}(\rb,t)$, which includes the density perturbation caused by charge rearrangement due to electron thermodiffusion $n_{\rm e}=n_{\rm e}^0+\Delta n_{\rm e}$, where the uniform unperturbed electron density is $n_{\rm e}^0=\me E_{\rm F}/\pi\hbar^2$ in the 2DEG and $n_{\rm e}^0= E_{\rm F}^2/\pi\hbar^2 v_{\rm F}^2$ in graphene (see Sec.\ \ref{secS1}). Because $n_{\rm e}^0\gg\Delta n_{\rm e}$ (see Fig.\ 4), we can safely use $n_{\rm e}^0$ and $E_{\rm F}$ to calculate the chemical potential $\mu(\rb,t,T)$. Nonetheless, the electric field $\Eb_{\rm Coul}$ generated by the nonuniform charge distribution should be calculated from $\Delta n_{\rm e}$.

The dynamics of the distribution function $f(\rb,\pb,t)$ is governed by the BTE, which, in the relaxation time approximation, can be written as
\begin{align}
\partial_t f+ \vb_\kb\cdot\nabla f+\Fb\cdot\nabla_{\pb} f=-\frac{(f-f^0)}{\tau(E_\kb)}, \nonumber
\end{align}
where the operator $\nabla$ without subscript it defined to act on space coordinates $\rb$, $\vb_\kb=\nabla_{\kb}E_\kb/\hbar$ is the electron group velocity, $\pb=\hbar\kb$ is the electron momentum, and $\Fb=-e\Eb_{\rm Coul}$ is the Lorentz force acting on the electrons (the contribution of magnetic fields is negligible). The gradients of $f$ in the real and momentum spaces can be explicitly found to be
\begin{align}
\nabla f&=\frac{\partial f}{\partial \mu} \nabla\mu    +    \frac{\partial f}{\partial T}  \nabla T
             = -\left(\nabla\mu + \frac{E_\kb - \mu}{T}\nabla T\right)\frac{\partial f}{\partial E_\kb}    ,\nonumber\\
\nabla_\pb f&=    \frac{\partial f}{\partial E_\kb} \vb_\kb,
\nonumber
\end{align}
which allows us to rewrite the BTE as
\begin{align}
\tau\partial_t f +(f-f^0)=\tau \vb_\kb \cdot \left( e  \Eb_{\rm Coul}+\nabla \mu + \frac{E - \mu}{T} \nabla T  \right) \frac{\partial f}{\partial E_\kb}.
\nonumber
\end{align}
Taking into account the contributions of multiple electron bands denoted by an index $\ell$, and considering that the electric current is determined by $\jb= -ge \sum_{\ell\kb} f(E_{\ell\kb})\vb_{\ell\kb} $, the integral of the above equation leads to
\begin{align}
&\mean{\tau\partial_t \jb}+\jb =  e^2\mathcal{J}_0 \Eb+e\mathcal{J}_0 \nabla \mu +\frac{e}{T}\mathcal{J}_1 \nabla T,  \label{je}
\end{align}
in terms of
\begin{align}
\mean{\tau\partial_t \jb}=   -ge \sum_{\ell\kb} \tau(E_{\ell\kb})f(E_{\ell\kb})\vb_{\ell\kb}\nonumber
\end{align}
and
\begin{align}
\mathcal{J}_m=&g  \sum_{\ell}\int \tau(E_{\ell\kb}) \frac{\dd\kb}{(2\pi)^2}  \vb_{\ell\kb}\otimes \vb_{\ell\kb}  \nonumber\\
          &\times (E_{\ell\kb}-\mu)^m  \left(-\frac{\partial f(E_{\ell\kb})}{\partial E_{\ell\kb}}\right)     ,\nonumber
\end{align}
where $g=g_{\rm s} g_{\rm v}$ accounts for spin degeneracy ($g_{\rm s}=2$), as well as valley degeneracy in graphene ($g_{\rm v}=2$); we have $g_{\rm v}=1$ for the 2DEG. At low temperatures, Eq.\ (\ref{je}) reduces to the Drude model because $\mean{\tau\partial_t \jb}\approx \tau(\mu)\partial_t \jb$. For isotropic systems, the integral of the dyadic term $\hat{\kb}\otimes\hat{\kb}$ over the azimuthal angle just contributes a factor of $\pi$. Finally, for both 2DEG and graphene, $\mathcal{J}_m$ simplifies to
\begin{align}
\mathcal{J}_m=\frac{1}{\pi\hbar^2}  \int  |E| \dd E   (E-\mu)^m \tau(E)
\left(-\frac{\partial f(E)}{\partial E}\right),
\nonumber
\end{align}
where the energy integration limits those mentioned in Sec.\ \ref{secS1}.

Equation\ (\ref{je}) can be recast in the form
\begin{subequations}
\begin{align}
&\tau\partial_t \jb+\jb =  \sigma_0 \Eb_{\rm tot}, \\
&\Eb_{\rm tot} = \Eb_{\rm Coul}+\nabla\mu-S\nabla T,
\end{align}
\label{je2}
\end{subequations}
where $\sigma_0=e^2 \mathcal{J}_0$ is the DC electric conductivity, and $S=-\mathcal{J}_1/\mathcal{J}_0 T e$ is the Seebeck coefficient (generally negative for electrons and positive for holes).

\section{Energy-dependent relaxation times}

\subsection{Scattering by charge impurities}

In the Born approximation, the scattering rate of an electron by the Coulomb potential associated with ion impurities in a 2D system is given by \cite{A06_2,HAS07}
\begin{align}
\gamma_{\kb\rightarrow\kb'}=n_{\rm i} \frac{2\pi}{\hbar} |\bra{\kb'}\frac{V(\qb)}{\epsilon(q,T)}\ket{\kb}|^2  \delta(E_\kb-E_{\kb'}),
\nonumber
\end{align}
where $n_{\rm i}$ is the area density of the said impurities, $\hbar\qb=\hbar(\kb'-\kb)$ is the change of electron momentum during scattering, and $\epsilon(q,T)$ is the screening function. We assume that ionic impurities are located on the plane of the 2D material and each of them has a charge $e$. The 2D Fourier component of the Coulomb potential $V(\qb)=v(q)\ee^{\ii\qb\cdot\rb}/q=2\pi e^2\ee^{\ii\qb\cdot\rb}/q$, evaluated at a wave vector $\qb=\kb'-\kb$, is screened by the 2D material as $V(\qb)/\epsilon(q,T)$, which provides the true potential leading to electron scattering in the above scattering matrix element.

The relaxation time $\tau(E)$ in the BTE due to the scattering by charge impurities can be found by summing over all scattering channels according to
\begin{align}
\frac{1}{\tau(E_\kb,T)}=\sum_{\kb'} \gamma_{\kb\rightarrow\kb'} [1-\cos(\theta_{\kb'\kb})]. \label{tauim}
\end{align}
We note that the factor $[1-\cos(\theta_{\kb'\kb})]$ properly introduces a dependence on scattering angle $\theta_{\kb'\kb}=\theta_{\kb'}-\theta_{\kb}$. For charge-impurity scattering, the temperature dependence of $\tau(E_\kb,T)$ is incorporated through the screening function $\epsilon(q,T)$.

We can apply the above approach to both 2DEG and graphene. The electron wave function for 2DEG is simply $\ket{\kb}=\ee^{\ii \kb\cdot\rb}/\sqrt{A}$, while for graphene the we need to use a Dirac spinor $\ket{\kb}=\ee^{\ii \kb\cdot\rb}  [ \ee^{-\ii\theta_\kb/2},\pm\ee^{\ii\theta_\kb/2}]^{\rm T}/\sqrt{2A}$, where $A$ is the surface normalization area. With the factor $\delta(E_\kb-E_{\kb'})$ rewritten as $\delta(k- k')(\me/\hbar^2 k)$ for 2DEG and $\delta(k- k')/\vF\hbar$ for graphene, we can carry out the $k'$ integral and only the angular integral remains in Eq.\ (\ref{tauim}). The relaxation time for both 2DEG and graphene finally reduces to
\begin{align}
\frac{1}{\tau(E_k,T)}=\frac{2 \pi^2 n_{\rm i} e^4 }{g_{\rm v}\hbar } N(E_k) \int_0^{2\pi} \frac{1}{|q\epsilon(q,T)|^2}    F [1-\cos(\theta)],  \nonumber
\end{align}
where $q=2k\sin(\theta/2)$, $F=1$ for 2DEG and $F=[1+\cos(\theta)]/2$ for graphene, and the valley degeneracy is $g_{\rm v}=1$ for 2DEG and $g_{\rm v}=2$ graphene.

The screening function used above can be written as $\epsilon(q,T)=1+v\chi(q,T)$ in terms of the temperature- and chemical-potential-dependent susceptibility $\chi(q,T,\mu)$. At zero temperature, the susceptibilities of 2DEG and graphene are given by \cite{AFS1982,WSS06,HD07}
\begin{flalign}
&\frac{\chi(q, 0, E_{\rm F})}{N(E_{\rm F})} =1- \Theta(1-x) \nonumber\\
&\times \left\{
\begin{aligned}
&\sqrt{1-x^2},               \quad (\text{for 2DEG})   \\
&\frac{1}{2}\sqrt{1-x^2}   + \frac{1}{2x}\cos^{-1}(x) ,    \quad (\text{for graphene})  \label{chi0}
\end{aligned}
\right.
\end{flalign}
where $x=2\kF/q$. The first term on the right-hand side of this expression is all that remains if $q<2\kF$ in both 2DEG and graphene; it corresponds to the Thomas-Fermi screening, which produces $\epsilon(q,T=0)=1+ k_{\rm TF}/q$, where $k_{\rm TF}=2\pi e^2N(E_{\rm F})$ is the Thomas-Fermi wave vector and $N(E_{\rm F}$ is the electron density of states at the Fermi level (see Sec.\ \ref{secS1}).

The susceptibility at finite temperature can be found from the zero temperature expression in Eq.\ (\ref{chi0}) according to \cite{M1978}
\begin{align}
\chi(q, T, \mu)=\frac{1}{4\kB T}\int_0^\infty \; \dd E \chi(q, 0, E)\, {\rm sech}^2 \left(\frac{\mu-E}{2\kB T}\right). \label{chiT}
\end{align}
In our study, the chemical potential $\mu(T)$ at the studied temperature $T$ is first found numerically following the approach discussed in Sec.\ \ref{secS1}, and $\chi(q, T, \mu)$ at this target temperature is calculated using Eq.\ (\ref{chiT}). We note that $\chi(q, 0, E)$ in Eq.\ (\ref{chiT}) is evaluated at arbitrary $E$, so the variable $x$ in Eq.\ (\ref{chi0}) should be determined as $x=2k_E/q$, where $\hbar^2 k_E^2/2\me=E$ for the 2DEG and $\vF\hbar k_E=E$ for graphene.

\subsection{Acoustic phonon scattering}

We consider the displacement field operator associated with a phonon of frequency $\omega_q$, wave vector of $\qb$, and polarization vector $\ub_\qb$:
\begin{align}
\Ub=\sqrt{\frac{\hbar}{2\omega_q\rho A}} ( \ub_\qb \hat{b} \ee^{\ii \qb\cdot \rb}+\ub_\qb^\dagger \hat{b}^\dagger \ee^{-\ii \qb\cdot \rb}),  \label{disp}
\end{align}
where $\rho$ is the mass area density, and $\hat{b}$ and $\hat{b}^\dagger$ are phonon creation and annihilation operators. The corresponding strain tensor is defined as $\bar{\bar{E}}=\nabla(\Ub+\Ub^T)/2$.

The interaction Hamiltonian obtained from linear deformation potential theory is $\hat{H}_{\rm I}= \bar{\bar{D}}_{\rm ac}\nabla \Ub$, where $\bar{\bar{D}}$ is the deformation potential \cite{HD08,HK09}. Considering the smallness of the phonon wave vector compared to that of the electrons, both the 2DEG and graphene can be regarded as isotropic, so the contributions of the transverse and longitudinal fields to the energy can be decoupled as
\begin{align}
\hat{H}_{{\rm I},j} = \sqrt{\frac{\hbar}{2\omega_0\rho A}}\ii q D_{{\rm ac},j}(\hat{b} \ee^{\ii \qb\cdot \rb}-  \hat{b}^\dagger \ee^{-\ii \qb\cdot \rb})  \quad (j=T,L) .\nonumber
\end{align}
The following results are equivalent for transverse and longitudinal components, so we disregard the subscript $j$ and sum over both contributions in the final result.

For the quantum states combining the electron and phonon subspaces $\ket{\kb,\{{n}\}_\qb}$, the transition matrix elements of the interaction Hamiltonian become
\begin{align}
|\bra{\kb',\{{n-1}\}_\qb}\hat{H}_{\rm I}\ket{\kb,\{{n}\}_\qb}|^2  &=   n    \frac{\hbar q^2 D^2}{2\omega_0\rho A}    |\bra{\kb'} \ee^{\ii \qb\cdot \rb} \ket{\kb}|^2, \nonumber\\
|\bra{\kb',\{{n+1}\}_\qb}\hat{H}_{\rm I}\ket{\kb,\{{n}\}_\qb}|^2  &=  (n+1)         \nonumber\\
                &\times\frac{\hbar q^2 D^2}{2\omega_0\rho A}    |\bra{\kb'} \ee^{\ii \qb\cdot \rb} \ket{\kb}|^2, \nonumber
\end{align}
where $n$ denotes phonon occupation numbers. Now, using Fermi's golden rule, the rate of electron scattering from states $\ket{\kb}$ to $\ket{\kb'}$ is given by
\begin{align}
\gamma_{\kb\rightarrow\kb'}=&\sum_{\qb,j}\frac{\pi q^2 D^2}{\omega_q\rho A} F \{ n_T(\omega_q)    \delta(E_{k'}-E_k-\hbar\omega_q) \nonumber\\
    &+ [n_T(\omega_q)+1] \delta(E_{k'}-E_k+\hbar\omega_q) \}\delta(\kb'-\kb-\qb), \nonumber
\end{align}
where we have performed a sum over all possible photon number states $\ket{n}_\qb$, leading to the Bose-Einstein distribution $n_T$ evaluated at the lattice temperature $T$ (assumed to be locally equal to the electron temperature), and we include a sum over phonon modes labeled by $\qb$ and $j$. The first and second terms inside the curly brackets account for photon absorption and emission, respectively. With this scattering rate, the resulting relaxation time associated with acoustic phonon scattering can be found in a way similar to Eq.\ (\ref{tauim}). Considering that low-energy phonons dominate the scattering process, the equipartition approximation $n_T(\omega_q)\approx \kB T/\hbar\omega_q$ can be adopted, and the two energy $\delta$ functions reduce to $\delta(E_{k'}-E_k)$. By further writing $\delta(E_{k'}-E_k)$ in momentum space as done above for charge-impurity scattering in the 2DEG and graphene, and then performing the integral over $\kb'$, we finally find the relaxation time associated with acoustic phonon scattering:
\begin{align}
\frac{1}{\tau(E,T)}= \alpha \frac{\pi D_{\rm ac}^2}{4\hbar\rho v_{\rm ac}^2} \kB T N(E),    \nonumber
\end{align}
where have taken into account contributions of both longitudinal and transverse acoustic phonons, and we define $2/v_{\rm ac}^2=1/v_{\rm ac,T}^2+1/v_{\rm ac,L}^2$ in terms of the sound velocity of transverse and longitudinal phonons, $v_{\rm ac,T}$ and $v_{\rm ac,L}$, respectively. Also, $\alpha=4$ for 2DEG and $\alpha=1$ for graphene.

\subsection{Optical phonon scattering}

The interaction Hamiltonian obtained from linear deformation potential theory for optical phonons is $\hat{H}_I= \bar{\bar{D}}_{\rm op}\cdot \Ub$. We note a difference in units between $\bar{\bar{D}}_{\rm op}$ and $\bar{\bar{D}}_{\rm ac}$ \cite{HK09}. The displacement field of optical phonons is also quantized by analogy to Eq.\ (\ref{disp}), with $\omega_q$ substituted by a constant frequency $\omega_0$ at $q\rightarrow 0$ for both transverse and longitudinal optical phonons, in contrast to the low energy acoustic phonon dispersion $\omega_\qb=v_{\rm ac} q$. The Hamiltonian of interaction between electrons and optical phonons is thus given by
\begin{align}
\hat{H}_{I}= \sqrt{\frac{2\hbar}{\omega_0\rho A}} D_{\rm op}(\hat{b} \ee^{\ii \qb\cdot \rb}+  \hat{b}^\dagger \ee^{-\ii \qb\cdot \rb})   ,\nonumber
\end{align}
where a factor of 2 is introduced to take into account the contributions of both transverse and longitudinal phonons.

Following a similar procedure as for acoustic phonons, we find the scattering rate of electrons by optical phonons to be
\begin{align}
\gamma_{\kb\rightarrow\kb'}=\frac{2\pi D^2}{\omega_0\rho A} F \{& n_T(\omega_0)    \delta(E_{k'}-E_k-\hbar\omega_0)   \nonumber\\
                                                                            & + [n_T(\omega_0)+1] \delta(E_{k'}-E_k+\hbar\omega_0)    \}.\nonumber
\end{align}
With this scattering rate, the relaxation time due to optical phonon scattering can be found by analogy to Eq.\ (\ref{tauim}) in the equipartition approximation as
\begin{align}
\frac{1}{\tau(E,T)}= \alpha \frac{\pi D_{\rm op}^2}{8\rho \omega_0} \{& n_T(\omega_0)  N(E+\hbar\omega_0)  \nonumber\\
                                                                      &+\left[n_T(\omega_0)+1\right]  N(E-\hbar\omega_0) \}. \nonumber
\end{align}

\section{2D Plasmon generation by electron thermodiffusion}

\begin{figure}
\begin{centering}
\includegraphics[width=0.5\textwidth]{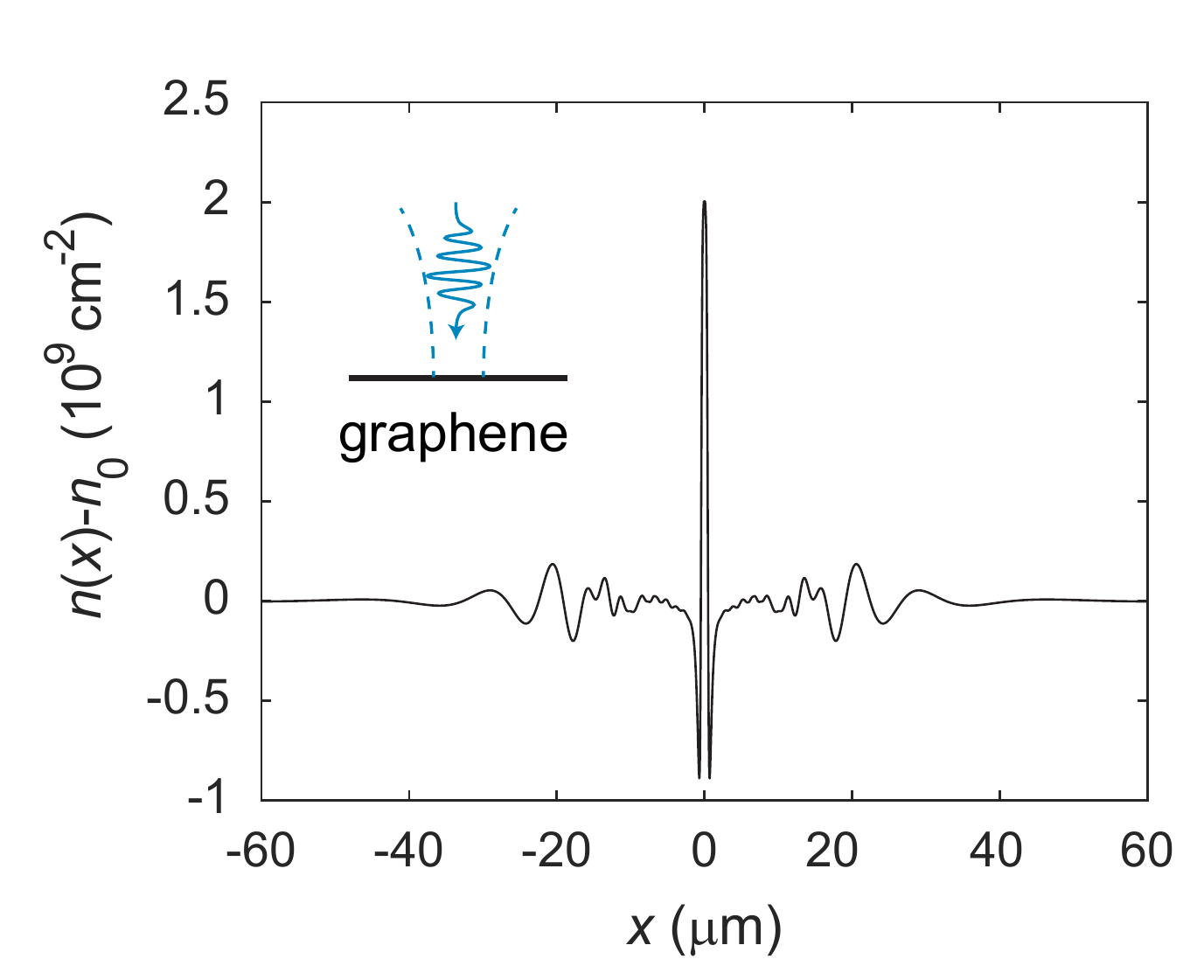}
\par\end{centering}
\caption{A snapshot of the electron density in a graphene layer heated by a focused laser pulse. Two chirped plasmon waves are observed, which propagate out of the heated region.}
\label{FigS2}
\end{figure}

Figure\ 4 in the main text shows that the rearrangement of electrons in a 2DEG or graphene layer eventually reaches a steady configuration under continuous heating by a focused CW laser beam. If we now consider the dynamical establishment of such nonuniform electron distribution, the emergence of the density perturbance starting from the homogeneous background in fact involves charge oscillations that can spread out as propagating plasmons.

The time-dependent evolution of the electron density under a dynamical temperature change is governed by Eqs.\ (\ref{je2}), together with the continuity equation $e\partial_t n(\rb)=\nabla\cdot\jb$. In the simulation presented below for graphene, these two equations are solved numerically using a real-space discretization combined with iteration in the time domain. The effect of laser-pulse heating can be modeled by a temperature evolution $T(x,t)=(T_{\rm max}-300[K])\ee^{-x^2/w^2}/[1+\ee^{-(t-t_0)/\Delta t}]+300[K]$, where the spatial distribution and temporal evolution are described by the Gaussian and logistic functions, respectively. This assumed evolution is realistic because thermal conduction is a much slower process compared to plasmon propagation, so the electrons are suddenly heated by a laser pulse to a certain temperature that is then maintained during the time period in which plasmons evolve in this study.

Figure\ \ref{FigS2} shows a snapshot at time at $t=3\,$ns of the resulting graphene electron density distribution excited by the temperature distribution defined above, where we assume the parameters $E_{\rm F}=0.2\,$eV, $w=500\,$nm, $T_{\rm max}=2000\,$K, and $t_0=\Delta t=50\,$fs. In addition to the strong electron rearrangement around the laser heated region that will be eventually relaxed to a steady electron distribution as shown in Figure\ 4, we also observe that two plasmon wave pulses carried by the charges oscillations propagate to $~20\,\mu$m away from the heating source. These two plasmon waves are chirped pulses, because the Fourier components of different frequencies propagate at different phase velocities due to the dispersion of the plasmon mode. This all-optical excitation approach of highly confined plasmons is feasible because the waist and duration of the incident laser pulse nearly match the wavelength and optical period of the graphene plasmons, respectively. In consequence, the phase-matching condition universally required to generate plasmons is naturally satisfied. Considering that the parameters of the temperature evolution function assumed above can be adjusted by changing the waist $w$ and duration $\Delta t$ of the incident laser pulse, the effect thus provides a feasible and tunable method to generate propagating plasmons in extended monolyaer graphene or even ultraconfined acoustic plasmons in double-layer graphene or metal-graphene structures \cite{LGA17}.

\bibliographystyle{apsrev}

\end{document}